\documentclass{amsart}
\usepackage{amsaddr}
\usepackage{graphicx}
\usepackage{epstopdf}
\usepackage{tabularx}
\usepackage{subfig}
\usepackage{amssymb}
\usepackage{amsthm}
\usepackage{amsmath}
\usepackage{algorithm} 
\usepackage{cite}

\theoremstyle{definition}

\theoremstyle{remark}

\numberwithin{equation}{section}

\newcommand{\bg}{\mathbf{g}}
\newcommand{\bL}{\boldsymbol{\Lambda}}
\newcommand{\bV}{\mathbf{V}}
\newcommand{\bv}{\mathbf{v}}
\newcommand{\bS}{\mathbf{S}}
\newcommand{\bqc}{\mathbf{q}^\mathrm{c}}
\newcommand{\bqs}{\mathbf{q}^\mathrm{s}}

\begin{document}

\title[Symmetrized Bingham Parameter Estimation]{Symmetrized Bingham Distribution for Representing Texture: Parameter estimation with respect to crystal and sample symmetries}

\author{Stephen R. Niezgoda}
\address{Department of Materials Science and Engineering, Department of Mechanical and Aerospace Engineering, The Ohio State University, Columbus, Ohio 43210}
%\curraddr{Department of Mathematics and Statistics,
%Case Western Reserve University, Cleveland, Ohio 43403}
\email{niezgoda.6@osu.edu}
\thanks{This work is supported by DARPA YFA Grant No. D15AP00103 "Computational Design Tools for Quantifying Uncertainty Due to Material Variability"}

\author{Eric A. Manguson}
\address{Department of Materials Science and Engineering, The Ohio State University, Columbus, Ohio 43210}

\author{Jared Glover}
\address{CapSen Robotics, Pittsburgh, Pennsylvania, 15206}
%    General info
%\subjclass[2000]{Primary 54C40, 14E20; Secondary 46E25, 20C20}

\date{\today}


%\keywords{Differential geometry, algebraic geometry}

\begin{abstract}
The quaternion Bingham distribution has been used to model preferred crystallographic orientation, or crystallographic texture, in polycrystalline materials in the materials science and geological communities. A primary difficulty in applying the Bingham distribution has been the lack of an efficient method for fitting the distribution parameters with respect to the materials underlying crystallographic symmetry or any statistical sample symmetry due to processing. In this paper we present a symmetrized distribution, based on the quaternion Bingham, which can account for any general combination of crystallographic or sample symmetries. We also introduce a numerical scheme for estimating the parameters of the symmetrized distribution based on the well known expectation maximization algorithm. %A simple example of the approach for fitting the symmetrized Bingham distribution for a material with a cubic crystal symmetry (Laue class $m\overline{3}m$) and orthotropic sample symmetry is also presented.
\end{abstract}

\maketitle

\section{Introduction}
\label{sec:intro}

The link between anisotropic properties and response in polycrystalline materials and crystallographic preferred orientations or texture has long been recognized \cite{bunge1982,Wenk1985,engler2010,wenk2004,bunge1987,wenk2002}.  The relative volume fraction of crystal orientations within a polycrystal is encoded by the orientation distribution function (ODF) \cite{bunge1982}. Macrotexture techniques, such as bulk X-ray or neutron diffraction, characterize the average texture over many grains and does not provide any spatial information concerning the arrangement of cystallites in a sample \cite{brokmeier1997,van1994}. Microtexture techniques on the other hand, including electron backscatter diffraction \cite{adams1993,dingley1997} or high energy X-ray diffraction microscopy \cite{hefferan2010,lienert2011} , directly and discretely measure local orientations and produce 2D or 3D spatially resolved orientation maps of a much smaller number of grains. 

Recently, Niezgoda and Glover proposed a novel approach to microtexture ODF estimation based on the idea that discrete orientation data are fundamentally different than bulk diffraction pole figure data \cite{niezgoda2013}. Rather than building up the ODF from the superposition of the contribution of individual orientations \cite{bunge1982, engler2010}, they posed the question \emph{``Can we find the ODF which was the most likely to give rise to the measured orientation data?"} Their approach consisted of three main components 1) The modeling of ODFs as a mixture of simpler distributions such as the Von Mises Fisher or Bingham distributions \cite{niezgoda2013}, 2) a minimum message length criterion to determine the ``optimum" number of components in the mixture model to avoid under/overfitting and 3) an unsupervised learning algorithm for fitting the parameters of the mixture model that maximizes the probability of measuring the set of discrete orientations. The critical weakness of the approach was a lack of a computationally efficient means of fitting the Bingham distribution parameters with respect to crystallographic and sample symmetry, and the case studies presented by Niezgoda and Glover were limited to the triclinic crystal symmetry ($C_1$ point group) \cite{niezgoda2013}. In this work, we introduce a symmetrized Bingham distribution,  describe an iterative approach to obtain expectation maximization maximum likelihood (EM-ML) estimates for the symmetrized Bingham parameters, and demonstrate the approach for fitting texture components in materials with cubic crystallographic and orthotropic statistical or sample symmetry. 

\section{Symmetrized Bingham Distribution}
\label{sec:method}

\subsection{The Quaternion Bingham Distribution}
\label{sec:bingham}
The Bingham distribution is an antipodally symmetric distribution defined on the unit hypersphere $\mathbb{S}^d \subset \mathbb{R}^{d+1}$ \cite{bingham1974}. Points on $\mathbb{S}^3$ are represented by the set of unit quaternions $U$, which is isomorphic to special unitary group of degree 2, $SU(2)$, and is a double covering group of the rotation group SO(3).  The Bingham distribution on $\mathbb{S}^3$, or the quaternion Bingham, is a convenient probability distribution to model texture and can represent common texture components including fibers, sheets, uniform, or anisotropic spreads around individual orientations \cite{schaeben1996,kunze2004,siemes2000,gorelova2014quantifying,bachman2010}. While the Bingham distribution is very flexible, there are ODFs for which a Bingham model is not suitable, such as the ‘‘cone fiber’’ textures explored in \cite{grewen1955,matthies1988}. As with all statistical modeling, it is important to test for goodness of fit, as even ``simple" texture components such as fibers and unimodal orientation spreads may be shaped in such a way that they are not be well represented by a Bingham model.

The  probability density function (p.d.f) for the quaternion Bingham is given by 
\begin{equation}
\label{eq:bing}
p(\bg;\bL,\bV)=\frac{1}{F(\bL)}\exp{\sum_{i=1}^4 \lambda_i(\bv_i \cdot \bg)^2}
\end{equation} 
where $\mathbf{g}$ is a unit quaternion representing an orientation, $\bL$ is a 4-vector of concentration parameters $\lambda_i$, and $F$ is a normalization constant. $\bV$ is a $4\times4$ matrix the columns of which, $\mathbf{v}_i$, are orthogonal unit quaternions representing the principal directions of the distribution. The dot operator $(\cdot)$ denotes the quaternion inner product. The concentration parameters, $\bL$, are unique only up to an additive constant. For this work we choose the convention $\lambda_1\leq \lambda_2 \leq \lambda_3 \leq \lambda_4 =0$ to resolve the ambiguity. The concentration parameters determine the sharpness of the distribution along the associated principal direction with $\bL=[0,0,0,0]$ corresponding with the uniform distribution. The primary difficulty in working with the Bingham distribution is the computation of the normalization constant $F(\bL)$ which is a generalized hypergeometric function with matrix argument. For fast processing the authors have precomputed $F(\bL)$ to a lookup table for a discrete grid of $\bL$ values. Interpolation is then used to quickly estimate normalization constants on the fly for arbitrary $\bL$ values \cite{niezgoda2013,glover2011}.

The maximum likelihood estimators (MLE) for the quaternion Bingham parameters, denoted $\hat{\bL}$ and $\hat{\bV}$, are straightforward to calculate. Given a set of $N$ discrete orientations, $\mathcal{G}=\lbrace \bg^{(1)},\dots ,\bg^{(N)}\rbrace$, the scatter matrix 
\begin{equation}
\label{eq:scatter}
\bS=\frac{1}{N}\sum_{i=1}^n \bg^{(i)} {\bg^{(i)}}^T=E\lbrace \bg \bg^T\rbrace
\end{equation}
is a sufficient statistic to calculate both $\hat{\bL}$ and $\hat{\bV}$. $\hat{\bV}$ is found by performing an eigenvalue decomposition of $\bS$. The MLE mode of the distribution is the eigenvector of $\bS$ with the largest eigenvalue and the columns of $\hat{\bV}$ are the eigenvectors corresponding to the 2nd, 3rd, and 4th eigenvalues. $\hat{\bL}$ is found by setting the partial derivatives of the log-likelihood function, $\log{p(\mathcal{G};\bL,\bV)}=\sum_{i=1}^N \log{p(\bg^{(i)};\bL,\bV) }$ with respect to the components of $\bL$ to zero yielding
\begin{equation}
\label{eq:MLestimate}
\frac{1}{F(\bL)}\frac{\partial F(\bL)}{\partial \lambda_j}=\bv_j^T \bS \bv_j.
\end{equation}
The values of the derivatives are precomputed and stored as a lookup table. Since the tables for $F$ and $\nabla F$ are indexed by $\bL$, a kD-tree is used to efficiently find the nearest neighbors for a computed $\nabla F / F$ to compute $\hat{\bL}$ by interpolation.

\subsection{Symmetry Group Invariant Bingham Distribution}
In order to practically apply the Bingham distribution to ODF representation we need to account for the underlying crystallographic and sample symmetries of the material and processing operations. Let $\mathcal{Q}^\mathrm{c}=\lbrace  \bqc_1,\dots \bqc_M \rbrace$ denote a group whose elements transform one orientation to a crystallography equivalent one. %$\mathcal{Q}^\mathrm{c}$ is a subgroup of the crystallographic point group excluding the reflection operations; as those transform a right handed coordinate system to a left handed coordinate system.% Formally $\mathcal{Q}^\mathrm{c}$ is a finite topological group which acts on $SO(3)$, $\bqc_i :SO(3) \rightarrow SO(3)$. The group operation is left multiplication which defines the composition of rotation $\bqc_i \ast \bqc_j$ as rotation $\bqc_j$ followed by $\bqc_i$. For parameterization by quaternions the $\ast$ operator denotes quaternion multiplication. 
If $\bqc \in\mathcal{Q}^\mathrm{c}$ then any function where $f(\bqc \ast \bg)=f(\bg)$ is said to be invariant under $\mathcal{Q}^\mathrm{c}$. A material sample can also contain statistical symmetry due to processing. The classic example is a statistical two-fold rotation axis about the rolling, transverse, and normal directions of a rolled plate resulting in orthotropic sample symmetry \cite{kocks1998}. The sample symmetry group $\mathcal{Q}^\mathrm{s}=\lbrace \bqs_1,\dots,\bqs_P \rbrace$ is defined in an identical way to the crystal symmetry group except the group operation is right multiplication.% and invariance under $\mathcal{Q}^\mathrm{s}$ implies $f(\bg \ast \bqs)=f(\bg)$. 
For materials with both symmetries any function of $\bg$ must be invariant under both symmetries as \cite{bunge1982}, 
\begin{equation}
f(\bqc \ast \bg \ast \bqs)=f(\bg).
\end{equation}

Chen et al. recently derived the form that all probability density functions must have in order to be invariant under spherical symmetry groups \cite{chen2015}. Here we trivially extend their result to include sample symmetry and state that the density function $p:SO(3) \rightarrow \mathbb{R}$ is jointly invariant under $\mathcal{Q}^\mathrm{c}$ and $\mathcal{Q}^\mathrm{s}$ if and only if
\begin{equation}
\label{eq:sym_p}
p(\bg;\boldsymbol{\Theta})=\frac{1}{M}\frac{1}{P}\sum_{i=1}^M \sum_{j=1}^P p(\bqc_i \ast \bg \ast \bqs_j;\boldsymbol{\Theta})
\end{equation}
where $\boldsymbol\Theta$ are the parameters of the probability density. Eq. \ref{eq:sym_p} states that any probability density $p(\bg)$ over the orientations  which is invariant to crystallographic and sample symmetry can be represented as a finite mixture with equal weights of the rotated density under the combined crystallographic and sample symmetry groups actions. Readers interested in a formal proof are referred to \cite{chen2015}.

Applying eq. \ref{eq:sym_p} allows us to directly write the p.d.f for the symmetrized Bingham as\footnote{It should be noted that the symmetrized Bingham distribution is not formally a Bingham distribution or even in the family of related exponential distributions. As discussed by \cite{kunze2005}, the summation over symmetry elements spoils the exponential form. We will keep the name symmetrized Bingham for convenience, as it aids understanding of how it is constructed} 
\begin{eqnarray}
\label{eq:sym_bingham_a}
p(\bg;\mathcal{Q},\bL,\bV)=\frac{1}{MP}\sum_{i=1}^M \sum_{j=1}^P p(\bqc_i \ast \bg \ast \bqs; \bL,\bV)\\
\label{eq:sym_bingham_b}
=\frac{1}{MP}\sum_{i=1}^M \sum_{j=1}^P p(\bg; \bL,\mathbf{Q}_i^\mathrm{c} \bV \mathbf{Q}_j^\mathrm{s}) = \frac{1}{MP}\sum_{i=1}^M \sum_{j=1}^P p(\bg; \bL,\bV_{ij})\\
\label{eq:sym_bingham_c}
=\frac{1}{MP}\frac{1}{F(\bL)}\sum_{i=1}^M \sum_{j=1}^P \left[ \exp{\sum_{k=1}^4 \lambda_i\left( \left[\bV_{ij} \right]_k\cdot \bg\right)^2}\right]
\end{eqnarray}
where $\mathcal{Q}$ denotes the symmetry groups, $\mathbf{Q}_r$ denotes the quaternionic matrix where the product $\mathbf{Q}_r \bV$ is equivalent to applying rotation $\mathbf{q}_r$ to each column of $\bV$, and $[\bV_{ij}]_k$ denotes the $k$th column of $V_{ij}$. Going from eq. \ref{eq:sym_bingham_a} to eq. \ref{eq:sym_bingham_b} requires the application of the inner quaternion product $\mathbf{v}_k \cdot \bg$ in  eq. \ref{eq:bing} and the observation that the inverse of symmetry elements must also be elements of the symmetry group. EQ. \ref{eq:sym_bingham_c} states that the symmetrized Bingham distribution is a finite mixture of the standard quaternion Bingham distributions, with each component a) having equal weight b) having principal directions rotated by $\mathbf{Q}_i^\mathrm{c} \bV \mathbf{Q}_j^\mathrm{s}=\bV_{ij}$ and c) having the same concentration parameters $\bL$. A similar weighted mixture mixture was defined in \cite{gorelova2014quantifying} and termed the Pseudo-Bingham distribution.

Given set $\mathcal{G}$ containing $N$ orientation measurements, the data log-likelihood is given by
\begin{eqnarray}
\label{eq:log_like}
\ln{p(\mathcal{G};\bL,\bV)}&=-N \ln{\left(MP F(\bL) \right)}+\\
&\sum_{n=1}^N \ln{
\sum_{i=1}^M \sum_{j=1}^P \exp{\left( \sum_{k=1}^4 \lambda_k \left( [\bV_{ij}]_k \cdot \bg_n\right)^2\right)}} \nonumber
\end{eqnarray}
While the symmetrized Bingham has an intuitive interpretation, the form of eq. \ref{eq:log_like}, specifically the sum over the symmetry elements, destroys the simple solution of the ML estimate of the Bingham Parameters described in Sec. \ref{sec:bingham}. It is possible to find the ML estimates by optimization, as expression of the gradients of the log-likelihood with respect to $\bL$ and $\bV$ is straightforward. However this approach was found to be highly inefficient due to the number of independent parameters, non-linear constraints, and numerical non-smoothness of the derivatives due to the approximation of the series expansion for $F(\bL)$ and interpolation of $F$ and $\partial F/ \partial \bL$ from the lookup tables. Instead we adopt an iterative Expectation Maximization - Maximum Likelihood (EM-ML) approach, for parameter estimation \cite{moon1996}.

\section{EM-ML Algorithm for Parameter Estimation}
\subsection{Mathematical Formulation}
As mentioned in Sec. \ref{sec:bingham}, $\bS$ is a sufficient statistic for parameter estimation for the standard Bingham distribution. Eq. \ref{eq:sym_bingham_b} shows that the symmetrized Bingham is finite mixture of rotated standard Bingham distributions. If $\bS$ could be calculated for the symmetrized case the standard ML estimates $\hat{\bL}$ and $\hat{\bV}$ could be trivially computed. Consider a set of $N$ discrete orientations, $\mathcal{G}=\lbrace \bg^{(1)},\dots ,\bg^{(N)}\rbrace$ which was generated by sampling a symmetrized Bingham distribution. In order to calculate $\bS$ we are missing information. The complete dataset would also contain a label which identifies which of the $MP$ rotated components of the mixture generated each sample \cite{niezgoda2013,chen2015,moon1996}. The EM-ML algorithm seeks the Bingham ODF which maximizes the probability of measuring the orientation data by a) estimating the labels given an estimate of the Bingham parameters (E-step) then b) using this new estimate of the labels to update the Bingham parameters (M-step). The algorithm is described more formally below.

For compactness let $\Theta=\lbrace{\mathcal{Q},\bL,\bV}\rbrace$ denote the complete set of parameters necessary to specify the symmetrized Bingham $p(\mathbf{g};\Theta)$. Further let $\theta_{ij}=\lbrace \bL,\bV_{ij}=\mathbf{Q}^c_i \bV \mathbf{Q}^s_j\rbrace$ denote the set of parameters necessary to specify an individual rotated component of the mixture, $p(\mathbf{g};\theta_{ij})$. 
Define a set of binary label vectors $\mathcal{Z}=\left[\mathbf{z}^{(1)},\dots,\mathbf{z}^{(N)} \right]$, where the elements of each vector $z^{(n)}_{ij}$ take the value 1 if $\bg^{(n)}$ was generated by $p(\mathbf{g};\theta_{ij})$ and 0 otherwise. If $\mathcal{Z}$ could be measured, then computing the scatter matrix and finding the ML estimate of the parameters $\hat{\Theta}$ would be trivial.

The complete data log-likelihood is given by
\begin{equation}
\label{eq:data_like}
\log{p(\mathcal{G},\mathcal{Z};\Theta)} = \frac{1}{MP}\sum_{n=1}^N \sum_{i=1}^M \sum_{j=1}^P
z^{(n)}_{ij} \log{p(g^{(n)};\theta_{ij})}.
\end{equation}
An ideal algorithm would maximize eq. \ref{eq:data_like} directly. However optimization of functions of binary variables (i.e. $\mathcal{Z}$) is problematic \cite{khanesar2007}. Instead we define the conditional expectation,
$\mathcal{W}=E\left[\mathcal{Z}\vert \mathcal{G},\hat{{\Theta}}\right]$. $w^{(n)}_{ij}$ gives the probability that $z^{(n)}_{ij}=1$ or equivalently the probability that orientation $\bg^{(n)}$ was generated by the rotated Bingham $p(\mathbf{g};\theta_{ij})$. By substituting $\mathcal{W}$ into eq. \ref{eq:data_like} we can use Bayes rule to update the probabilities of mixture assignments. In the EM literature this is termed the $Q$ function. If $\hat{\Theta}$ represents a current estimate of the parameters then
\begin{eqnarray}
\label{eq:Qa}
Q(\Theta \vert \hat{\Theta}) &= E \left[\log{p(\mathcal{G},\mathcal{Z} \vert \Theta)} \vert \mathcal{G},\hat{\Theta}\right]\\
\label{eq:Qb}
&= \log{p(\mathcal{G},\mathcal{W} \vert \Theta)}\nonumber
\end{eqnarray}

The E-step consists of using Bayes rule to update $\mathcal{W}$ as 
\begin{eqnarray}
\label{eq:E}
w^{(n)}_{ij}&=E[z^{(n)}_{ij}\vert \mathcal{G},\hat{\Theta}]=\mathcal{P}[z^{(n)}_{ij}=1 \vert \bg^{(n)},\hat{\Theta}]\\
\label{eq:bayes}
&= \frac{(MP)^{-1}p(\bg^{(n)};\hat{\theta}_{ij})}
{\sum_{s,t}(MP)^{-1}p(\bg^{(n)};\hat{\theta}_{st})}.\nonumber
\end{eqnarray}
Then in the M-step the parameter estimates $\hat{\Theta}$ are updated to maximize eq. \ref{eq:Qa}. The ML estimate for $Q(\Theta \vert \hat{\Theta})$ can then be derived, by setting the derivatives with respect to the parameters to zero and solving. The ML estimates take exactly the same form as those given for the standard Bingham in sec. \ref{sec:bingham} if the scatter matrix is replaced by $\mathbf{S}(\mathcal{Q})$ the symmetrized scatter matrix
\begin{equation}
\label{eq:M}
\mathbf{S}(\mathcal{Q})=\frac{1}{N}\sum_{n=1}^N \sum_{i=1}^M \sum_{j=1}^P w^{(n)}_{ij} \bg^{(n)}_{ij} {\bg_{ij}^{(n)}}^T
\end{equation}
where $\bg^{(n)}_{ij}=({\bqs_i})^{-1} \ast \bg^{(n)} \ast ({\bqc_j})^{-1}$. The EM-ML algorithm alternates between eq. \ref{eq:E} and eq. \ref{eq:M} until convergence is reached. For this work convergence was defined as the change in eq. \ref{eq:Qa} between iterations was less than some small value , $\Delta Q(\Theta \vert \hat{\Theta})/n \leq \delta$. 

\subsection{Example}
As a demonstration we apply the algorithm to the case of texture estimation in a cubic-orthorhombic material, meaning that each orientation has 96 symmetric equivalents (the composition of 24 from cubic crystallographic Laue class $m\bar{3} m$ and 4 from sample symmetry group). $n=1000$ i.i.d. sample orientations, and the stopping criterion $\delta=10^{-5}$ was used. In order to define a ``ground-truth'' for comparison, $n$ quaternions were sampled from a non-symmetrized Bingham distribution with known parameters. The ML estimates of the quaternion Bingham parameters (Eqs. \ref{eq:scatter} and \ref{eq:MLestimate}) were computed from the samples and the resulting fit Bingham was used as the ground truth.  The sampled quaternions were each replaced with a randomly chosen symmetrically equivalent orientation to produce $\mathcal{G}$. In this way we ensured direct comparison between the EM-ML process to fit the symmetrized Bingham and the ML estimates of the standard quaternion Bingham distribution using equivalent initial data. The quality of fit was evaluated by computing the data log likelihood (Eq. \ref{eq:log_like}) and the texture entropy for each ODF as well as the integrated error between the fit ODF and the ground truth.

Figure 1 shows the results of this procedure. For convenience the ODFs are plotted with respect to the Bunge-Euler angles at constant $\phi_2$ sections, for interpretation of the ODF images see \cite{bunge1982,kocks1998}. 1000 samples were drawn from an anisotropic Bingham distribution with $\Lambda=[-25,-20,-15]$
and $V$ chosen as a random orthogonal matrix. As expected the fit ODF is virtually identical to the ground truth ODF. Specific details of the goodness of the fit are given in Table 1. It is interesting to note that the likelihood of the symmetrized fit is slightly larger than the ground truth fit. Additionally the progression of the algorithm, shown by the evolution of $ Q(\Theta \vert \hat{\Theta})$ with each iteration, is highlighted in Fig. 2.

\begin{table}
\caption{Comparison of ground truth versus fit symmetrized Bingham ODF.}
\begin{tabular}{lcccc}      % Alignment for each cell: l=left, c=center, r=right
{} & $\Lambda$ & Log Likelihood & Integrated Error & Entropy\\
\hline
Initial Target & [-25.00 -20.00 -15.00] & 181.63 & 0.0281 & -0.1926\\
Ground Turth & [-24.98 -19.11 -15.46] & 182.23 & 0.000 & -0.2058\\
Fit Symmetric & [-24.76 -18.79 -15.80] & 183.45 & 0.0096 & -0.2027\\
\end{tabular}
\end{table}

\begin{figure}
\captionsetup{type=figure}
\subfloat[]{
\includegraphics[scale=0.25,keepaspectratio=true]{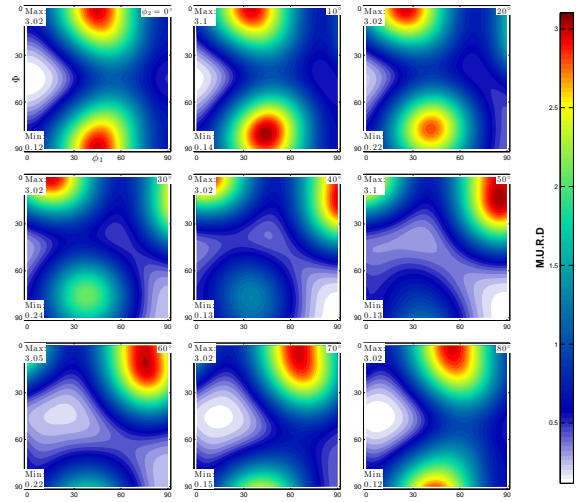}
\label{fig:fit_example_1_a}
} \\
\subfloat[]{
\includegraphics[scale=0.25,keepaspectratio=true]{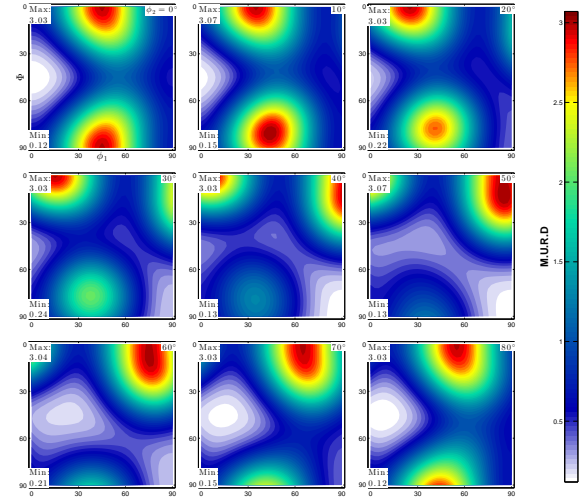}
\label{fig:fit_example_1_b}
}
\caption{Comparison of the ground truth ODF (a) against the fit ODF for a material with cubic crystal symmetry and orthotropic sample symmetry. For convenience the ODFs are plotted as $\phi_2$ sections of the Bunge-Euler angles \cite{bunge1982}, as is routinely done in the quantitative texture analysis literature.} 
\label{fig:fit_example_0}
\end{figure}

\begin{figure}
\captionsetup{type=figure}
\includegraphics[scale=0.35]{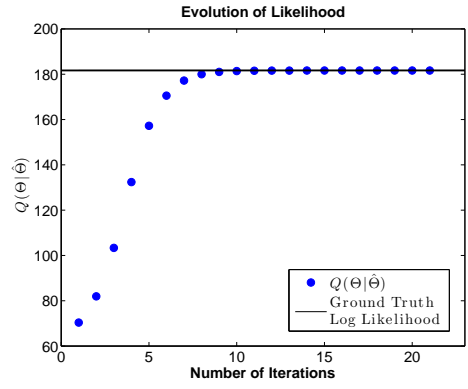}
\caption{Comparison of ground truth versus fit symmetrized Bingham ODF}
\label{fig:likelihood_evolution}
\end{figure}

\clearpage
\bibliographystyle{unsrt}
\bibliography{sym_bingham}

\begin{thebibliography}{10}

\bibitem{bunge1982}
H.-J. Bunge.
\newblock {\em Texture analysis in materials science : mathematical methods}.
\newblock Butterworths, 1982.

\bibitem{Wenk1985}
H.R. Wenk.
\newblock {\em Preferred orientations in deformed metals and rocks: an
  introduction to modern texture analysis}.
\newblock Academic Press, Orlando, 1985.

\bibitem{engler2010}
Olaf. Engler and V.~Randle.
\newblock {\em Introduction to texture analysis : macrotexture, microtexture,
  and orientation mapping}.
\newblock CRC Press, 2010.

\bibitem{wenk2004}
H-R Wenk and P~{Van Houtte}.
\newblock Texture and anisotropy.
\newblock {\em Reports on Progress in Physics}, 67(8):1367, 2004.

\bibitem{bunge1987}
H.J. Bunge.
\newblock Three-dimensional texture analysis.
\newblock {\em International Materials Reviews}, 32(1):265--291, 1987.

\bibitem{wenk2002}
Hans-Rudolph Wenk.
\newblock Texture and anisotropy.
\newblock {\em Reviews in Mineralogy and Geochemistry}, 51(1):291--329, 2002.

\bibitem{brokmeier1997}
HG~Brokmeier.
\newblock Neutron diffraction texture analysis.
\newblock {\em Physica B: Condensed Matter}, 234:977--979, 1997.

\bibitem{van1994}
B.A. Van Der~Pluijm, N.C. Ho, and D.R. Peacor.
\newblock High-resolution x-ray texture goniometry.
\newblock {\em Journal of Structural Geology}, 16(7):1029--1032, 1994.

\bibitem{adams1993}
Brent Adams, Stuart Wright, and Karsten Kunze.
\newblock Orientation imaging: The emergence of a new microscopy.
\newblock {\em Metallurgical and Materials Transactions A}, 24:819--831, 1993.

\bibitem{dingley1997}
D.~J. Dingley and D.~P. Field.
\newblock Electron backscatter diffraction and orientation imaging microscopy.
\newblock {\em Materials Science and Technology}, 13(1):69--78, 1997.

\bibitem{hefferan2010}
CM~Hefferan, SF~Li, J.~Lind, U.~Lienert, AD~Rollett, P.~Wynblatt, and RM~Suter.
\newblock Statistics of high purity nickel microstructure from high energy
  x-ray diffraction microscopy.
\newblock {\em Computers, Materials, \& Continua}, 14(3):209--220, 2010.

\bibitem{lienert2011}
U~Lienert, SF~Li, CM~Hefferan, J~Lind, RM~Suter, JV~Bernier, NR~Barton,
  MC~Brandes, MJ~Mills, MP~Miller, et~al.
\newblock High-energy diffraction microscopy at the advanced photon source.
\newblock {\em Jom}, 63(7):70--77, 2011.

\bibitem{niezgoda2013}
Stephen~R Niezgoda and Jared Glover.
\newblock Unsupervised learning for efficient texture estimation from limited
  discrete orientation data.
\newblock {\em Metallurgical and Materials Transactions A}, 44(11):4891--4905,
  2013.

\bibitem{bingham1974}
C.~Bingham.
\newblock An antipodally symmetric distribution on the sphere.
\newblock {\em The Annals of Statistics}, pages 1201--1225, 1974.

\bibitem{schaeben1996}
H.~Schaeben.
\newblock {Texture Approximation or Texture Modelling with Components
  Represented by the von Mises{--}Fisher Matrix Distribution on {\it SO}(3) and
  the Bingham Distribution on {\it S}${\sp {4+}}{\sb +}$}.
\newblock {\em Journal of Applied Crystallography}, 29(5):516--525, Oct 1996.

\bibitem{kunze2004}
Karsten Kunze and Helmut Schaeben.
\newblock The bingham distribution of quaternions and its spherical radon
  transform in texture analysis.
\newblock {\em Mathematical Geology}, 36:917--943, 2004.

\bibitem{siemes2000}
Heinrich Siemes, Helmut Schaeben, Carlos~A. Rosi{\`e}re, and Horst Quade.
\newblock Crystallographic and magnetic preferred orientation of hematite in
  banded iron ores.
\newblock {\em Journal of Structural Geology}, 22(11�12):1747 -- 1759, 2000.

\bibitem{gorelova2014quantifying}
S~Gorelova, H~Schaeben, and R~Kawalla.
\newblock Quantifying texture evolution during hot rolling of magnesium twin
  roll cast strip.
\newblock {\em Materials Science and Engineering: A}, 602:134--142, 2014.

\bibitem{bachman2010}
F.~Bachman, R.~Hielscher, and H.~Schaeben.
\newblock Texture analysis with mtex- free and open source software toolboc.
\newblock {\em Solid State Phenomena}, 160:63--68, 2010.

\bibitem{grewen1955}
Johanna Grewen and G~Wassermann.
\newblock {\"U}ber die idealen orientierungen einer walztextur.
\newblock {\em Acta Metallurgica}, 3(4):354--360, 1955.

\bibitem{matthies1988}
S~Matthies, K~Helming, T~Steinkopff, and K~Kunze.
\newblock Standard distributions for the case of fibre textures.
\newblock {\em physica status solidi (b)}, 150(1):K1--K5, 1988.

\bibitem{glover2011}
J.~Glover, G.~Bradski, and R.B. Rusu.
\newblock Monte carlo pose estimation with quaternion kernels and the bingham
  distribution.
\newblock In {\em Proceedings of Robotics: Science and Systems VII}, Los
  Angeles, California, USA, June 2011.

\bibitem{kocks1998}
U.~F. Kocks, C.~N. Tom{\'{e}}, and Hans-Rudolf Wenk.
\newblock {\em Texture and anisotropy : preferred orientations in polycrystals
  and their effect on materials properties}.
\newblock Cambridge University Press, 1998.

\bibitem{chen2015}
Yu-Hui Chen, Dennis Wei, Gregory Newstadt, Marc DeGraef, Jeffrey Simmons, and
  Alfred Hero.
\newblock Parameter estimation in spherical symmetry groups.
\newblock {\em Signal Processing Letters, IEEE}, 22(8):1152--1155, 2015.

\bibitem{kunze2005}
K~Kunze and H~Schaeben.
\newblock Ideal patterns of preferred crystallographic orientation and their
  representation by the von mises fisher matrix or bingham quaternion
  distribution.
\newblock In {\em Proceedings of ICOTOM14, Materials Science Forum}, volume
  495, page 497, 2005.

\bibitem{moon1996}
T.K. Moon.
\newblock The expectation-maximization algorithm.
\newblock {\em Signal Processing Magazine, IEEE}, 13(6):47--60, 1996.

\bibitem{khanesar2007}
Mojtaba~Ahmadieh Khanesar, Mohammad Teshnehlab, and Mahdi~Aliyari Shoorehdeli.
\newblock A novel binary particle swarm optimization.
\newblock In {\em Control \& Automation, 2007. MED'07. Mediterranean Conference
  on}, pages 1--6. IEEE, 2007.

\end{thebibliography}

\end{document}